\documentclass[aps,prd,superscriptaddress,nofootinbib,amsmath,amsfonts,preprintnumbers,groupedaddress,showpacs,10pt,english]{revtex4-1}
\usepackage{amsmath}
\usepackage{amssymb}
\usepackage{babel}
\usepackage{wrapfig}
\usepackage{cancel}

\usepackage{relsize,exscale}
\makeatletter

\usepackage{array,multirow,graphicx}
\usepackage{dcolumn}
\usepackage{newlfont}
\usepackage{bm}
\usepackage[colorlinks,citecolor=blue,urlcolor=blue,linkcolor=blue]{hyperref}
\usepackage[figtopcap]{subfigure}
\usepackage{color}

\newcommand{\pont}{{\,^\ast\!}R\,R}

\newcommand\be{\begin{equation}}
\newcommand\ba{\begin{eqnarray}}
\newcommand\ee{\end{equation}}
\newcommand\ea{\end{eqnarray}}

\usepackage{scalerel}
\usepackage{tikz}
\usetikzlibrary{svg.path}
\definecolor{orcidlogocol}{HTML}{A6CE39}
\tikzset{
  orcidlogo/.pic={
    \fill[orcidlogocol] svg{M256,128c0,70.7-57.3,128-128,128C57.3,256,0,198.7,0,128C0,57.3,57.3,0,128,0C198.7,0,256,57.3,256,128z};
    \fill[white] svg{M86.3,186.2H70.9V79.1h15.4v48.4V186.2z}
                 svg{M108.9,79.1h41.6c39.6,0,57,28.3,57,53.6c0,27.5-21.5,53.6-56.8,53.6h-41.8V79.1z M124.3,172.4h24.5c34.9,0,42.9-26.5,42.9-39.7c0-21.5-13.7-39.7-43.7-39.7h-23.7V172.4z}
                 svg{M88.7,56.8c0,5.5-4.5,10.1-10.1,10.1c-5.6,0-10.1-4.6-10.1-10.1c0-5.6,4.5-10.1,10.1-10.1C84.2,46.7,88.7,51.3,88.7,56.8z};}}
\newcommand\orcid[1]{\href{https://orcid.org/#1}{\mbox{\scalerel*{
\begin{tikzpicture}[yscale=-1,transform shape]
\pic{orcidlogo};
\end{tikzpicture}
}{|}}}}
\begin{document}

\date{\today}
\title{Slow Kerr-NUT black hole solution in dynamical Chern-Simons modified gravity}

\author{G.~G.~L.~Nashed~\orcid{0000-0001-5544-1119}}
\email{nashed@bue.edu.eg}
\affiliation {Centre for Theoretical Physics, The British University, P.O. Box
43, El Sherouk City, Cairo 11837, Egypt}

\author{Kazuharu Bamba}%
\email{bamba@sss.fukushima-u.ac.jp}
\affiliation{Faculty of Symbiotic Systems Science,
Fukushima University, Fukushima 960-1296, Japan}

\begin{abstract}

The slow rotation of Kerr-NUT spacetime is explored by taking into account the linear form of rotation and NUT parameters in the dynamical Chern-Simon gravity theory, which can be formulated from a scalar field describing the background. We show that in the absence of the potential scalar field, the metric potential does not respect the effect of the NUT parameter, although the scalar field is affected by the rotation and NUT parameters. Consequently, unlike the gradually spinning black hole solution outlined in \cite{Alexander:2009tp}, the mixed component of the metric potential, encompassing both rotational and NUT parameters, doesn't make a contribution at the primary level of the initial perturbation.
\end{abstract}





\maketitle

\section{Introduction}
\label{intro}
Newman, Tamburino, and Unti (NUT) discovered one of the most intriguing GR solutions in 1963 \cite{Newman:1963yy}. This is a generalization of the Schwarzschild solution, which solves the Einstein vacuum field equations by including an extra parameter, the NUT charge $n$, in addition to the mass parameter M. It is commonly used to describe a gravitational dyon with both ordinary and magnetic mass. The NUT charge $n$ impacts in the same way that electric and magnetic charges do in Maxwell theory \cite{Plebanski:1975xfb}.  The NUT charge has numerous interpretations and meanings \cite{Lynden-Bell:1996dpw}, and there are numerous explanations for the physical origin of NUT-charged spacetimes \cite{Griffiths:2005qp}.

Alternatively, within the scientific community, we can demonstrate a model as presented in \cite{Jackiw:2003pm}, wherein the Einstein-Hilbert action is coupled to the topological Pontryagin term through an auxiliary field \cite{Jackiw:2003pm}. The external field within this theory offers intriguing facets, including the fact that the Schwarzschild spacetime remains a solution in the modified framework. As a result, the fundamental predictions of the theory of general relativity formulated by Albert Einstein remain unchanged. Additionally, the conceptual framework  anticipates the transmission of gravitational waves at the speed of light, denoted by $c$, albeit with varying strengths, thereby introducing a departure from spatial reflection symmetry. Furthermore, the discourse outlined in \cite{Jackiw:2003pm} was conducted within the purview of the Lagrangian framework. This implies that any modification could potentially impact the canonical structure of the initial theory or even result in changes to the count of degrees of freedom. Hence, it becomes imperative to engage in a canonical analysis to ascertain the potential alterations in the canonical framework that might arise within the modified theory. In CS theory, one has two alternative scenarios, referred to as  non-dynamical and dynamical approaches. In the dCS, the pseudo-scalar field verifies its field equations. However, in the non-dynamics case, the scalar field is explained as being entirely  external quantity. Both scenarios offer modified equations of motion compared with Einstein's GR. In a non-dynamical framework, the effects coming from the expression for CS curvature which depend on C-tensor, that is know as the Cotton tensor \cite{Jackiw:2003pm,Alexander:2009tp}. Moreover, the amended equation of motion satisfies the Pontryagin constraint, ${\,^\ast\!}RR= 0$, which ensures the diffeomorphism invariance of the CS theory. Furthermore, the situation is radically changing in the dCS frame. This is due to two facts: Firstly, since the dynamical variable, $\varrho(x)$ yields an extra effect on the metric potential via the stress-energy tensor, besides the C-tensor. The second reason is  due to the scalar field, which verifies the dynamical field equation instead of the restriction given by  Pontryagin equation.

The importance of dCS theory in dealing with the problem raised in \cite{Johannsen:2010xs} can then be estimated given the Pontryagin density's symmetry properties under parity transformations. The condition that the dCS correction maintains parity implies that, in accordance with its hypothetical string theory, the scalar coupling should be mediated by a pseudo-scalar field. This suggests that any parity-even solutions, like Schwarzschild BH and any spherically symmetric metrics, are not affected by CS corrections, which appear only in the violation of parity like Kerr or any rotating solutions \cite{Yunes:2009hc,Nashed:2023qjm,Nashed:2023yjt,Nashed:2023qnw,Nashed:2022kes}. In another meanings, the CS expression can spontaneously create Kerr shifts and presents theoretically reasonable grounds \cite{Johannsen:2010xs}. Many novel BH solutions in the frame of dCS have been derived in the scientific society, like rotating  spacetimes of G\"{o}del-type \cite{Porfirio:2016nzr,Porfirio:2016ssx,Agudelo:2016pic,Altschul:2021rog} or  Einstein-Gauss-Bonnet  theory with dilaton \cite{Kanti:1995vq,Kanti:1997br,Kleihaus:2011tg,Ayzenberg:2014aka,Maselli:2015tta,Kleihaus:2015aje,Okounkova:2019zep,Cano:2019ore,Delgado:2020rev,Pierini:2021jxd}. The noteworthy effect of the CS expression in such BH solutions became critical to find a large category of full causal BH solutions different from Einstein's GR theory. Moreover, the basic properties of the Lorentzian NUT solution in dCS theory was discussed in \cite{Brihaye:2016lsx}. The perturbative charged rotating AdS black hole solutions for Einstein-Maxwell CS theory for arbitrary values of the coupling is studied in \cite{Mir:2016dio}. The regime of validity of the perturbation scheme in the frame of dCS is analyzed in \cite{Stein:2014xba}.

Moreover, the importance of dCS (dynamical Chern-Simons) within the realm of modified gravitational theories is greatly emphasized by diverse rationales arising from various physical foundations, where the existence of the CS term appears to be prevalent \cite{Alexander:2009tp}. Furthermore, the importance of dCS in the domain of modified gravitational theories is significantly accentuated by a variety of rationales originating from diverse physical foundations, wherein the presence of the CS expression seems to be widespread \cite{Alexander:2009tp}. Especially in physics, for example, to remove the anomaly, the Lagrangian should include a CS-like counter term proportional to the Pontryagin density.  Counter terms like type can be generated  in string theory to generate low-energy effective string models using the Green-Schwarz mechanism \cite{Smith:2007jm, Adak:2008yg}. Numerous parallels can be identified in comparison to the techniques of loop quantum gravity \cite{Ashtekar:1988sw}, where the CS term emerges to resolve the ambiguity of the Immirzi field and the chiral anomaly of fermions \cite{Perez:2005pm,Freidel:2005sn,Date:2008rb,Mercuri:2009zi,Mercuri:2009vk}. Moreover, the CS theory might contribute to devising novel strategies for detecting violations of the local Lorentz/CPT symmetry within the gravitational field, potentially leading to the acquisition of fresh observational data in the forthcoming years.  Indeed, the consequences of parity violations due to CS terms are widely recognized across various contexts \cite{Jackiw:2003pm,Martin-Ruiz:2017cjt,Nojiri:2019nar,Nojiri:2020pqr,Nojiri:2020pqr}, including applications to the baryon asymmetry problem \cite{Alexander:2004us,Garcia-Bellido:2003wva,Alexander:2004xd}, as well as considerations related to cosmic microwave background polarization \cite{Alexander:2006mt,Lue:1998mq,Bartolo:2018elp,Bartolo:2017szm}.

The present work aims to study the slow of Kerr-NUT in the frame of dCS gravity and compare the output results with the results presented in \cite{Yunes:2007ss} to investigate the effect of the NUT parameter.

The structure of this investigation is outlined as follows:

In Section~\ref{ABC}, we provide a concise overview of the modified CS gravitational theory.

In Section~\ref{axisym}, we utilize the equations of motion derived from the gravitational theory of dCS to analyze the line element of a slowly rotating Kerr-Newman-Unti-Tamburino (Kerr-NUT) black hole. Our focus is particularly on cases with minor coupling constants in the Chern-Simons framework

We analyze the effective Newtonian potential of our solution in Section~\ref{sec4}.

The central aspect of this study is presented in Section~\ref{conclusions}.

Throughout this study, the following symbols are consistently utilized:

In a 4-dimensional spacetime, we adopt the signature $(-,+,+,+)$~\cite{Misner:1973prb}, and the round and square brackets denote symmetrization and antisymmetrization, respectively. The notation $\partial \varphi/\partial r=\partial_r\varphi=\varphi_{,r}$ signifies partial derivatives.

We employ geometrized units where $G=c=1$, and Einstein summation convention is applied.
\section{Chern-Simon modified gravitational theory}
\label{ABC}

We will  present in this section the topics which support a full  formulation of the CS-amendment  theory of gravitation and give some symbols
\cite{Alexander:2009tp}.

\subsection{Brief summary of CS  theory}
The Lagrangian of a CS theory of gravity takes the form:
\be
\label{CSaction}
L =L_{EH} + L_{CS} +  L_{\varrho} + L_{M}\,,
\ee
where $L_{EH}$ is the Einstein Hilbert Lagrangian that has the form:
\be
\label{EH-action}
L_{EH} = \kappa \int_{V} d^4x  \sqrt{-g}  R,
\ee
${L}_{CS}$ is the Chern-Simon Lagrangian defined as:
\be
\label{CS-action}
L_{CS}= \frac{\gamma}{4} \int_{V} d^4x  \sqrt{-g} \;
\varrho \; \pont\,,
\ee
$L_{\varrho}$ is the Lagrangian of the scalar field given by:
\be
\label{Theta-action}
L_{\varrho} = - \frac{\gamma_1 }{2} \int_{V} d^{4}x \sqrt{-g} \left[ g^{\alpha \beta}
\left(\nabla_{\alpha} \varrho\right) \left(\nabla_{\beta} \varrho\right) + 2 V(\varrho) \right], \quad
\ee
$L_{M}$ is the Lagrangian of the matter field given by:
\be
L_{M}= \int_{V} d^{4}x \sqrt{-g} {L}_{mat.}.
\ee

Subsequently, we introduce the ensuing symbols that will be utilized in the ongoing exploration: $\gamma$ and $\gamma_1$ representing dimensional constants are $\gamma$ and $\gamma_1$; $\nabla_{\alpha}$ stands for the  derivative; $R$ indicates the scalar of  Ricci tensor; and $g$ corresponds to the metric determinant; and $\kappa = \frac{1}{16 \pi G}$. Furthermore, the symbol $\pont$ designates the Pontryagin density, which assumes the following structure:
\be
\label{pontryagindef}
\pont= {\,^\ast\!}R^\alpha{}_{\beta}{}^{ \gamma \delta} R^{\beta}{}_{\alpha \gamma \delta}\,,
\ee
where ${\,^\ast\!}R^\alpha{}_{\beta}{}^{ \gamma \delta}$ refers to the dual Riemann-tensor  figured as:
\be
\label{Rdual}
{\,^\ast\!}R^\alpha{}_{\beta}{}^{ \gamma \delta}=\frac12 \epsilon^{ \gamma \delta \gamma_1 \delta_1}R^\alpha{}_{\beta \gamma_1 \delta_1}\,,
\ee
where $\epsilon^{ \gamma \delta \gamma_1 \delta_1}$ is a tensor that is entirely skew-symmetric, characterized by $\epsilon^{0123}$=-1.

The CS scalar field, $\varrho$, is the term that parameterizes the shift from the theory of GR. If $\varrho = constant$, then the Pontryagin density becomes the total derivative,  and we can recover Einstein's GR theory as:
\be
\nabla_\alpha \Upsilon^a = \frac{1}{2} \pont,
\label{eq:curr1}
\ee
where
\be
\Upsilon^\alpha =\epsilon^{\alpha \beta \gamma \delta} \Gamma^{\rho}_{{\beta} \epsilon} \left(\partial_{\gamma}\Gamma^{\epsilon}_{\delta \rho}+\frac{2}{3} \Gamma^{\epsilon}_{\gamma \gamma_1}\Gamma^{\gamma_1}_{\delta \rho}\right)\,,
\label{eq:curr2}
\ee
where  $\Gamma$  is the second kind Christoffel  symbols. By using  Eq. \eqref{eq:curr2} we can write $S_{\textrm{CS}}$ in the  form \cite{Yunes:2007ss}:
\be
\label{CS-action-K}
S_{{CS}} =  \gamma
\left( \varrho \; {\Upsilon}^{\alpha} \right)|_{\partial {{V}}}
-
 \frac{\gamma}{2} \int_{{{V}}} d^4x  \sqrt{-g} \;
\left(\nabla_{\alpha} \varrho \right) {\Upsilon}^{\alpha}\,.
\ee
The usual practice involves ignoring the initial term in Eq. \eqref{CS-action-K} since its effects are taken into account at the spacetime's surface~\cite{Grumiller:2008ie}, while the second term is recognized as the Chern-Simons expression.

The variation of the Lagrangian \eqref{CSaction} regarding the metric tensor, and the CS field yields the following equation of motions:
%
\ba
\label{eom}
&&R_{ab} + \frac{\sigma_1}{\kappa} C_{\alpha \beta} = \frac{1}{2 \kappa} \left(T_{\alpha \beta} - \frac{1}{2} g_{\alpha \beta} T \right),
\\
\label{eq:constraint}
&&\gamma_1 \; \square \varrho = \gamma_1 \; \frac{dV}{d\varrho} - \frac{\gamma}{4} \pont\,.
\ea
Here, $\square $ is the D'Alembertian operator and $R_{ab}$ is the tensor of Ricci. The expression $C_{ab}$ pertains to the C-tensor, which is characterized by the following definition:
\be
\label{Ctensor}
C^{\alpha \beta} = v_\gamma
\epsilon^{\gamma \gamma_1\gamma_2(\alpha}\nabla_{\gamma_2}R^{\beta){}_{\gamma_1}}+v_{\alpha_1\alpha_2}{\,^\ast\!}R^{\alpha_2(\alpha \beta)\gamma}\,,
\ee
with
\be
\label{v}
v_\alpha=\nabla_\alpha\varrho\,,\qquad
v_{\alpha \beta}=\nabla_\alpha\nabla_{\beta}\varrho\,.
\ee
In conclusion, the complete stress-energy tensor is formulated as:
\be\label{Tab-total}
T_{\alpha \beta} = T^{{mat}}_{\alpha \beta} + T_{\alpha \beta}^{\varrho},
\ee
where $T^{mat}_{\alpha \beta}$ is the matter sources (which we will omit it in this study)
\ba
\label{Tab-theta}
&&T_{\alpha \beta}^{\varrho}
=   \gamma_1  \left[  \left(\nabla_{\alpha} \varrho\right) \left(\nabla_{\beta} \varrho\right)
    - \frac{1}{2}  g_{\alpha \beta}\left(\nabla^{\alpha_1} \varrho\right) \left(\nabla^{\alpha_1} \varphi\right)
-  g_{\alpha \beta}  V(\varrho)  \right]\,.
\ea
Within the context of the gravitational CS theory, the robust equivalence principle, which states ($\nabla_{\beta} T^{\alpha \beta}_{{mat}} = 0$), is upheld under the condition that the scalar field $\varrho$ satisfies the field equations as stated in Eq.\eqref{eq:constraint}. This arises because when we differentiate Eq. \eqref{eom}, the initial term on the left-hand side becomes null as a consequence of the Bianchi identities. Nevertheless, the association between the second term and the Pontryagin density is established by the subsequent correlation:
\be
\label{nablaC}
\nabla_\alpha C^{\alpha \beta} = - \frac{1}{8} v^\beta \pont.
\ee
The satisfaction of Eq. \eqref{nablaC} gives  Eq.~\eqref{eq:constraint}.

To end this section, we will discuss the coupling constant's dimensions employed throughout this study and the scalar field $\varrho$. The determination of the units of one of
$(\gamma,\gamma_1,\varrho)$ will determine the other units.
Illustratively, when the CS field incorporates the unit $[\varrho] = l^{a}$, this results in $[\gamma] = l^{2 - a}$ and $[\gamma_1] = l^{-2a}$, where $l$ represents the unit of length.  The normal process which determines the CS scalar, $\varrho$, to be a dimensionless quantity, as normally done in the scalar-tensor theories, which requires  $[\gamma] = l^{2}$ and $\gamma_1$ to be dimensionless~\footnote{In this context, we adopt units such that $G = c = 1$, resulting in the action being measured in units of $l^{2}$. Consequently, if we use natural units where $G = 1 = c$, the action becomes dimensionless. In the scenario where $[\varrho] = l^{a}$, this leads to $[\gamma] = l^{-a}$ and $[\gamma_1] = l^{-2 s - 2}$.}. An alternative option is to consider $\gamma = \gamma_1$, effectively treating $S_{\varrho}$ and the action of ${CS}$ on the same level; this implies $[\varrho] = l^{-2}$. However, no particular framework compels us to adopt specific units for $\varrho$, hence we will keep these undetermined, as earlier investigations have led to distinct choices.
\section{Solution for a rotating-NUT BH in the context of dCS gravity}\label{axisym}

Next, our focus turns to examining rotating-NUT black holes within the framework of dCS gravity. Highlighting the importance, delving into the characteristics of stationary axisymmetric spacetime under the influence of dCS gravity, without resorting to simplifications in the analysis, poses a demanding and intricate task. Hence, we will employ two approximations to streamline the assessment process. Subsequently, our aim is to tackle the solution of the modified dCS equations of motion through the application of a second-order perturbation expansion methodology.
\subsection{The approximate process}
\label{approx}

Our approach will involve the utilization of two approximation techniques: The approximations involving gradual rotation and NUT parameter denoted by $a$ and $n$, as well as the scenario of small coupling.
The approach of small-coupling employs the modified Chern-Simons term as a slight departure from Einstein's General Relativity, allowing us to represent the metric decomposition (up to the second order) in the subsequent manner:
\ba\label{small-cou-exp0}
&&g_{\alpha \beta} = g_{\alpha \beta}^{(0)} + \zeta g^{(1)}_{\alpha \beta}(\varrho) + \zeta^{2} g^{(2)}_{\alpha \beta}(\varrho)\,. \nonumber\\
%
\ea
In this context, $g_{\alpha \beta}^{(0)}$ stands for the background metric that fulfills the field equations of General Relativity, exemplified by the Kerr-NUT metric. Nevertheless, $g_{\alpha \beta}^{(1)}(\psi)$ and $g_{\alpha \beta}^{(2)}(\varrho)$ represent the perturbations at first and second orders, respectively, originating from the modified Chern-Simons term. The parameter $\zeta$ corresponds to the level of approximation within the small-coupling framework.
Furthermore, the approximation involving gradual rotation and NUT parameter allows us to redefine both the background and $\zeta$-perturbations using the Kerr-NUT rotation parameter $a_{K}$ and the NUT parameter $n$ as power series. This leads to the following expressions for the metric background and metric perturbations:
\ba
\label{small-cou-exp}
g_{\alpha \beta}^{(0)} &=& \eta_1{_{\alpha \beta}^{(0,0)}} + \eta \; h_{\alpha \beta}^{(1,0)} + \eta^{2} h_{\alpha \beta}^{(2,0)},
\nonumber \\
\zeta g_{a{a_1}}^{(1)} &=& \zeta h_{\alpha \beta}^{(0,1)} + \zeta \eta \; h_{\alpha \beta}^{(1,1)} + \zeta \eta^{2} h_{\alpha \beta}^{(2,1)},
\nonumber \\
\zeta^{2} g_{\alpha \beta}^{(2)} &=& \zeta^{2} h_{\alpha \beta}^{(0,2)} + \zeta^{2} \eta \; h_{\alpha \beta}^{(1,2)} + \zeta^{2} \eta^{2} h_{\alpha \beta}^{(2,2)}\,.
\ea
Here, $\epsilon$ signifies the first order of the slow rotation-NUT expansion. It's important to highlight that the symbol $h^{\alpha \beta}_{m n}$ represents expressions of ${\cal{O}}(\alpha \beta)$, denoting an expression of ${\cal{O}}(\eta^{\alpha})$ and ${\cal{O}}(\zeta^{a_1})$.

To exemplify this concept, let's consider Eq.~\eqref{small-cou-exp}, where $\eta_1{{\alpha \beta}}^{(0,0)}$ and $\eta_1{{\alpha}}^{(0,0)}$ symbolize the reference metric when the rotation parameter is set to zero, i.e., $a_{K} = 0$. Conversely, $h_{a{a_1}}^{(1,0)}$, $h_{a{a_1}}^{(2,0)}$, $\zeta_{a}^{(1,0)}$, and $\zeta_{a}^{(2,0)}$ indicate the perturbations at the first and second orders with respect to the reference metric and the charge associated with the angular momentum parameter.

By merging the earlier outlined approximation methods, we arrive at an expansion that depends on the two distinct parameters, $\zeta$ and $\eta$, culminating in the formulation of the second-order perturbation for both the metric and charge, which adopt the subsequent expressions:
\ba
&&g_{\alpha \beta} = \eta_1{_{\alpha \beta}^{(0,0)}} + \eta h_{\alpha \beta}^{(1,0)} + \zeta h_{\alpha \beta}^{(0,1)} + \eta \zeta h_{\alpha \beta}^{(1,1)} + \eta^{2} h_{\alpha \beta}^{(2,0)} + \zeta^{2} h_{\alpha \beta}^{(0,2)}\nonumber\\
%
\ea

The elements associated with the leading order are labeled as ${\cal{O}}(1,0)$ or ${\cal{O}}(0,1)$, while the subsequent order components include ${\cal{O}}(2,0)$, ${\cal{O}}(0,2)$, or ${\cal{O}}(1,1)$.

In the context of such investigation, the process of gradual rotation involves expanding the Kerr-NUT parameter, indicated as $a_{K}$. As a result, its expansion becomes dimensionless, specifically represented by $a_{K}/M$. Consequently, the resulting equations, upon multiplication by $\eta^n$, exhibit an order of ${\cal{O}}\left((a_{K}/M)^n\right)$.
\subsection{The solution for a slowly rotating Kerr-NUT black hole}\label{slow-rot}

The expansion for slow rotation and NUT parameter can be constructed utilizing the Hartle-Thorne approximation~\cite{Thorne:1984mz,Hartle:1968si}, where the line element can be expressed in terms of the following parameterization:
\ba
\label{slow-rot-ds2}
&&ds^{2} = -p  \left[1 + p_1(r,\theta)\right] dt^{2}
+ \frac{1}{p}  \left[1 + p_2(r,\theta)\right] dr^{2}
+ r^{2} \left[1 + p_3(r,\theta) \right] d\theta^{2}
+r^{2} \sin^{2}{\theta} \left[1 + p_4(r,\theta) \right] \left[ d\phi  - \upsilon(r,\theta) dt \right]^{2}\,,\nonumber\\
&&\ea
where $p$ is defined as  $p = 1 - \frac{2 M}{r}$ which is the Schwarzschild  solution and we consider here as the background metric, where $M$ signifies the mass  when the CS term is not considered.
In Eq.~\eqref{slow-rot-ds2}, we employ the coordinates $(t,r,\theta,\phi)$, and the perturbations in the metric are indicated by $p_1(r,\theta)$, $p_2(r,\theta)$, $p_3(r,\theta)$, $p_4(r,\theta)$, and $\upsilon(r,\theta)$.

Metric~\eqref{slow-rot-ds2} is rewrote similar to the one presented  in~\cite{Thorne:1984mz,Hartle:1968si} and up to  the second order expansion yields:
\ba \label{cons}
 p_1(r,\theta) &=& \eta \; p_1{_{(1,0)}} + \eta\;  \zeta \;  p_1{_{(1,1)}} + \eta^{2} \;  p_1{_{(2,0)}},
 \nonumber \\
 p_2(r,\theta) &=& \eta \; p_2{_{(1,0)}} + \eta\;  \zeta \;  p_2{_{(1,1)}} + \eta^{2} \;  p_2{_{(2,0)}},
 \nonumber \\
 p_3(r,\theta) &=& \eta\; p_3{_{(1,0)}} + \eta\;  \zeta \;  p_3{_{(1,1)}} + \eta^{2} \;  p_3{_{(2,0)}},
 \nonumber \\
 p_4(r,\theta) &=& \eta\; p_4{_{(1,0)}} + \eta\;  \zeta \;  p_4{_{(1,1)}} + \eta^{2} \;  p_4{_{(2,0)}}.
 \nonumber \\
  \upsilon(r,\theta) &=& \eta\; \upsilon_{(1,0)} + \eta \;  \zeta \; \upsilon_{(1,1)} + \eta^{2} \; \upsilon_{(2,0)}.
 \ea
Equations~\eqref{cons} have no terms of ${\cal{O}}(0,0)$ since such expression already exists in the  Schwarzschild structure of Eq.~\eqref{slow-rot-ds2} where we have bivariate the diagonal components of Kerr-NUT in terms of the rotation parameter, $a_K$, and the NUT parameter, $n$.
Furthermore, it is postulated that as the Kerr rotation parameter approaches zero ($a_{K}\rightarrow 0$), the resulting solution corresponds to the Schwarzschild NUT spacetime
that ensures all terms of ${\cal{O}}(0,a)$ are vanishing.
Hence, the CS expression must exhibit linearity concerning both the Kerr rotation and NUT parameters. By considering the bivariate nature of $a_K$ and $n$ within the Kerr-NUT of GR, the metric perturbation that can be expressed as being proportional to $\zeta^{0}$ up to the linear order:
\ba\label{0e}
p_1{_{(1,0)}} &=& p_2{_{(1,0)}} = p_3{_{(1,0)}} = p_4{_{(1,0)}} =0\,, \qquad
\upsilon_{(1,0)}=2\frac{ nr\cos\theta(1-\frac{2 M}{r})-Ma_K}{r}\,,
\ea
 and up to second order as:
\ba
p_1{_{(2,0)}} &=& -p_2{_{(2,0)}}=-\frac {{a_K}^{2}\,p\, \cos^2 \theta+2\,a_K\,n\,p\,  \cos\theta -{a_K}^{2} \left(1-\frac{2\,{n}^{2}}{{a_K}^{2}} \right)-\frac{2\,M{n}^{2}}{r} }{{r}^{2}\,p}\,,\nonumber \\
p_3{_{(2,0)}}& =& \frac{{a^2}_{K}\cos^{2}{\theta}+a_K\,n\cos\theta+n^2}{r^{2}}\,, \qquad \qquad \upsilon_{(2,0)} = 0\,,
\nonumber \\
p_4{_{(2,0)}} &=& \frac{a_K{}^2(2M\sin^4\theta-r\cos^2\theta)+2n\,a_K\cos\theta\sin^2\theta(r-4M)+n^2[5\sin^2\theta-4(r-2M\cos^2\theta)]}{r^2 \sin^{2}\theta}\,, \ea
which coincides with Kerr solution when $n=0$ \cite{Yunes:2007ss} and with Schwarzschild-NUT in the scenario where the rotation parameter becomes negligible, specifically when $a_K=0$, \cite{Nashed:2014zca,Nashed:2008qr,Nashed:2019uyi}.
All the fields are bivariate in terms of the rotation parameter $a_K$ and the NUT parameter, $n$ in addition to small-coupling approximation, incorporating CS  field. To obtain  the leading-order
of the scalar field  $\varrho$ we must calculate the evolution equation, Eq.~\eqref{eq:constraint}. Eq.~\eqref{eq:constraint} yields $\partial^{2} \varrho \sim (\gamma_1/\gamma) \pont$, from where one can show that
the Pontryagin density has a null value up to order  $a_{K}/M+n$ .
Hence, the initial order of the CS scalar must be $\varrho \sim (\sigma_1/\sigma) ((a_{K}/M)+n)$, which $\propto \eta$.
Additionally, the hypotheses  that the Schwarzschild  is the only solution up to the zero  NUT  and angular
momentum parameters yields $\varrho^{(0,a)} = 0$ to all $a$. In present research, we will try to derive slow rotating taking into account the effect of the NUT parameter. To accomplish this, we will examine two distinct scenarios: One where the rotation parameter becomes negligible, and another where neither the rotation nor the NUT parameters vanish.

\subsection{The case of $a_K$=0}
Through the utilization of Eq. \eqref{eq:constraint} on Eq. \eqref{slow-rot-ds2}, and employing Eq. \eqref{cons}, we derive:
\be
\label{th-ansatz}
\varrho = \eta \; \varrho^{(1,0)}(r,\theta) + \eta \;  \zeta \; \varrho^{(1,1)}(r,\theta) + \eta^{2} \;\varrho^{(2,0)}(r,\theta)\,.
\ee
We will now proceed to employ the method outlined earlier to solve the altered field equation presented above. By placing emphasis on the equation governing the evolution of the dCS scalar, with a specific focus on the ${\cal{O}}(1,0)$ order, we derive the equation describing the progression in the subsequent form:
\ba
\label{1st-eq1}
&&p \varrho^{(1,0)}_{,rr} + \frac{2}{r} \varrho^{(1,0)}_{,r} \left( 1 - \frac{M}{r} \right) + \frac{1}{r^{2}} \varrho^{(1,0)}_{,\theta \theta} + \frac{\cot{\theta}}{r^{2}} \varrho^{(1,0)}_{,\theta}
= \frac{48n\,M(r-3M)}{r^{7}} \frac{\gamma}{\gamma_1}  \,.
\ea
The solution to the aforementioned differential equation, denoted as \eqref{1st-eq1}, forms a linear amalgamation of the solutions to the homogeneous and particular equations: $\psi^{(1,0)} = \varrho^{(1,0)}{Hom.} + \varrho^{(1,0)}{Part.}$.
The homogeneous equation lends itself to separation:
\be\label{hom}
\varrho^{(1,0)}_{Hom.}(r,\theta) = \varrho(r)\varrho(\theta).
\ee
Eq. \eqref{hom} shows that the  differential equation \eqref{1st-eq1},  yields  ordinary differential equations of $\varrho(r)$ and $\varrho(\theta)$,
that their solutions  take the form:
\ba
\label{Hom-sol-1}
\varrho(r) &=&c_{1} {\mathrm H}\left[\left[\frac{\mathrm s}{2},\frac{\mathrm  s}{2}\right],{\mathrm s},\frac{2 M}{r} \right] r^{-{\mathrm  s}/2}+ c_{2}  {\mathrm H}\left[\left[\frac{{\mathrm s}_1}{2},\frac{{\mathrm s}_1}{2}\right],{\mathrm s}_1,\frac{2M}{r}\right] r^{-{\mathrm s}_1/2}\,,
\nonumber \\
\varrho(\theta) &=& c_{3} {\mathrm  L}[-s/2,\, \cos \theta] + c_{4} {{\mathrm L}_1}[-s/2,\, \cos\theta]\,.
\ea
Here, ${\mathrm H}(\cdots)$ represents generalized hypergeometric functions\footnote{{
The generalized hypergeometric function ${\textit HG}
\left( \left[n_1, n_2,\cdots, n_p \right], \left[d_1, d_2, \cdots, d_q \right], z\right)$ is
typically figured as,
\[
{\textit HG}
\left( \bm{n}, \bm{d}, z \right) = \sum_{k=0}^\infty \frac{\prod_{i=1}^p  \mathrm{PS} \left( n_i, k \right)}
{\prod_{j=1}^q \mathrm{PS} \left( d_j,k \right)} \frac{z^k}{k!}\, ,
\]
where $\bm{n} = \left[ n_1,n_2,\cdots, n_p \right]$, $\bm{d}= \left[ d_1,d_2,\cdots, d_q \right]$ and $\mathrm{PS}(n,k)$
represents the Pochhammer symbol, $\mathrm{PS}(n,k) \equiv \prod_{j=0}^{k-1} \left( n + j \right)$.
$H(\cdots)$'s in (\ref{Hom-sol-1}) correspond to $p=2$ and $q=1$.
}},
$L(\cdot)$ is the Legendre polynomial of the first kind\footnote{{
The definition pertaining to the first-type Legendre polynomial is as follows:
\[
{\textit L}(b,z)= {\textit HG}\left( [-b,b+1],[1],\frac{1}{2}(1-z) \right) \, .
\] }},
${{\textit L_1}}(\cdot)$ is the second-type Legendre polynomial\footnote{
The second-type Legendre polynomial is defined as,
\[
{\textit L_1}(b,z)=\frac{\sqrt{\pi} \Gamma(1+b) {\textit HG} \left( \left[ 1 + \frac{b}{2}, \frac{1}{2} + \frac{b}{2} \right], \left[ \frac{3}{2}+b \right],\frac{1}{z^2} \right)}
{2 z^{1+b} \Gamma \left( \frac{3}{2}+b \right)2^b}\, .
\] } and  $c_{i}$, $i=1 \cdots 4$ are constants and ${\mathrm  s}$ and ${\mathrm  s}_1$ have the following definitions:
\be
\label{tilde-alpha}
{\mathrm  s}= 1 - \sqrt{1 - 4 c_{5}}\,,
\qquad
{\mathrm  s}_1 = 1 + \sqrt{1 - 4 c_{5}},
\ee
where $c_{5}$ is another constant.

 We are going to investigate a comprehensive description of the solution for $\varrho^{(1,0)}$ as follows: to comprehend the physical implications of the constants of integration present within it. For this intention, we will explore the distant-field behavior of Eq. \eqref{Hom-sol-1}, i.e., when $r \gg M$, we get:
\be
\varrho(r) \sim c_{1} \left[ 1 + \frac{M}{2 r} {\mathrm  s} \right] r^{-{\mathrm  s}/2}
+  c_{2} \left[ 1 + \frac{M}{2 r} {\mathrm  s}_1\right] r^{-{\mathrm  s}_1/2} .
\ee
 Furthermore, we stipulate the requirement for the scalar field $\varrho$ to possess real values, resulting in $s \in \Re$ and $s_1 \in \Re$, thereby establishing $c_{5} < 1/4$. Additionally, assuming that $\psi$ possesses finite total energy far from the horizon necessitates a decay rate faster than $1/r$, implying ${\mathrm s} > 2$ and ${\mathrm s}1 > 2$. However, the first constraint does not hold true when $c{5} < 1/4$, leading to $c_{1} = 0$. Conversely, the second constraint results in $c_{5} < 0$. As a result, the requirements associated with ensuring definite total energy render $\psi$ not proportionate to $\ln(h)$.
From the  above discussion, we get:
\be \varrho^{(1,0)}_{Hom}=constant \,.\ee

With the homogeneous solution of Eq. \eqref{1st-eq1} in hand, we can proceed to deduce the specific solution, yielding:
\ba
\label{theta-sol-SR1114}
&&\varrho^{(1,0)}_{_{Part.}}(r,\theta)=2\left( r-M\right)\int \!{\frac {12\,M\alpha\,n\,\eta\,{r}^{2}+18\,{M}^{3} \gamma\,n\,\eta-32\,r{M}^{2}\gamma\,n\,\eta}{ \gamma_1\,{r}^{5} \left( 4\,{r}^{ 2}M-5\,r{M}^{2}-{r}^{3}+2\,{M}^{3} \right) }}{dr} \,.
\ea

The asymptotic form of Eq. \eqref{theta-sol-SR1114}  gives:
\ba
\label{theta-sol-SR1112}
&&\varrho^{(1,0)}_{_{Part.}}(r,\theta)\approx \frac{24\,n\,M\,\gamma}{5\gamma_1 r^4}\left(1+\frac{M}{9r}+\frac{21M^2}{126r^2}+\frac{5M^3}{14r^3}\right)+{\cal O}\left(\frac{1}{r^8}\right)\,.
\ea

Having effectively introduced the dCS field, we are now poised to embark on deriving the metric corrections associated with dCS.
It is important to highlight that the stress-energy tensor, as expressed in \eqref{Tab-theta}, corresponding to the dCS field, will follow the same equation of motion, as stated in \eqref{eom}, up to the ${\cal{O}}(2,1)$ order.As a consequence, its impact on the metric perturbation will be omitted. Under these circumstances, we can classify the altered Einstein equation of motions divided into two distinct categories:

The initial category comprises an independent set of the equations encompassing ${h_1}^{(1,1)}$, ${h_2}^{(1,1)}$, ${h_3}^{(1,1)}$, and ${h_4}^{(1,1)}$. These equations originate from the elements linked with $(t,t)$, $(r,r)$, $(r,\theta)$, $(\theta,\theta)$, and $(\phi,\phi)$.

The second category, arising from the revised Einstein equations, leads to a distinctive differential equation that governs $\omega^{(1,1)}$. This equation is responsible for controlling the $(t,\phi)$ component within the adapted Einstein field equations.

The initial category is independent of the dCS field, represented by $\psi$, resulting in the complete elimination of its contribution. Hence, our focus will be directed towards the second subset, specifically the $(t,\phi)$ component, which yields:
\ba
\label{V1}
p{r}^{3}\,\sin \theta  \,\cos \theta\, \upsilon_{rr} \left( r,\theta \right)+ r\sin \theta\,\cos\theta\, \upsilon_{\theta \theta} \left( r,\theta \right)-r \left( 1+\sin^2\theta\right)\, \upsilon_\theta \left( r, \theta \right) +2\,M\,\sin2\theta\,\upsilon \left( r, \theta \right) =0\,. \ea
The first information about the above differential equation is that it is a homogenous linear differential equation which means that it has only a homogenous solution, unlike the slow rotating solution.  Now let us discuss four different cases of

 Once more, the most general solution takes the structured as a linear amalgamation, comprising both a homogeneous solution and a designated particular solution known as $\upsilon \left( r, \theta \right)$:\\
 Case I:\\
 When  $\upsilon \left( r, \theta \right)=\upsilon( r)$ we get the solution of Eq. \eqref{V1} as
 \be \label{Or}
 \upsilon^{(1,1)}( r)=c_1\,p\,,\ee where $c_1$ is a constant of integration and we have put the other constant of this solution to be vanishing to get finite total energy of the BH. Eq. \eqref{Or} shows that we have no extra contribution of $\upsilon^{(1,1)}( r)$ because as Eq. \eqref{0e} shows that the leading order of ${\cal O}\left(\eta^0\right)$ will be ${\cal O}\left(\frac{1}{r}\right)$ which is the same as the leading order of $\upsilon^{(1,1)}( r)$.\\
 Case II:\\
 When  $\upsilon \left( r, \theta \right)=\upsilon( r)\cos\theta$ and by using the same procedure of Case I we get:
 \be \label{Oct}
 \upsilon^{(1,1)}( r)=r^2\left\{c_2\,r^{{\mathrm s}_2}\,p^{{\mathrm s}_2}\, {\mathrm H}\,\left[\left[{\mathrm s}_2,{\mathrm  s}_3\right],{\mathrm s}_4,\frac{-2 M}{r\,p} \right] +c_3\,r^{{\mathrm s}_5}\,p^{{\mathrm s}_5}\, {\mathrm H}\,\left[\left[{\mathrm s}_6,{\mathrm  s}_7\right],{\mathrm s}_8,\frac{-2 M}{r\,p} \right] \right\}\,,\ee where $c_2$ and $c_3$ are constants and ${\mathrm s}_2 \cdots {\mathrm s}_8 $ are defined as:
 \ba
&& {\mathrm s}_2=\frac{-3\cos\theta+\sqrt{9\cos^2\theta-8}}{2\cos\theta}\,,\qquad  {\mathrm s}_3=\frac{5\cos\theta-\sqrt{9\cos^2\theta-8}}{2\cos\theta}\,, \qquad {\mathrm s}_4=\frac{\cos\theta-\sqrt{9\cos^2\theta-8}}{\cos\theta}\,, \nonumber\\
 && {\mathrm s}_5=\frac{-3\cos\theta-\sqrt{9\cos^2\theta-8}}{2\cos\theta}\qquad {\mathrm s}_6=\frac{5\cos\theta+\sqrt{9\cos^2\theta-8}}{2\cos\theta}\,,\qquad  {\mathrm s}_7=\frac{3\cos\theta+\sqrt{9\cos^2\theta-8}}{2\cos\theta}\,, \nonumber\\
 &&{\mathrm s}_8=\frac{\cos\theta+\sqrt{9\cos^2\theta-8}}{\cos\theta}\,.\ea
 Eq. \eqref{Oct} have no finite value of energy therefore, the case $\upsilon \left( r, \theta \right)=\upsilon( r)\cos\theta$ yields a constant value of $\upsilon \left( r, \theta \right)$, i.e., $\upsilon \left( r, \theta \right)=constant$.\\
 Case III:\\
 When  $\upsilon \left( r, \theta \right)=\upsilon( r)\sin\theta$ and by using the same procedure of Case I we get:
 \be \label{Ost3}
 \upsilon^{(1,1)}( r)=p\left\{c_4\,{\mathrm H}\,\left[\left[{\mathrm s}_9,{\mathrm  s}_{10}\right],[-2],\frac{r}{2M} \right] +c_5r^3 {\mathrm H}\,\left[\left[{\mathrm s}_{11},{\mathrm  s}_{12}\right],[4],\frac{r}{2M} \right] \right\}\,,\ee where $c_4$ and $c_5$ are constants and ${\mathrm s}_9 \cdots {\mathrm s}_{12}$ are defined as:
 \ba
&& {\mathrm s}_9=\frac{-\sin\theta+\sqrt{13-9\cos^2\theta}}{2\sin\theta}\,,\qquad  {\mathrm s}_{10}=\frac{-\sin\theta-\sqrt{13-9\cos^2\theta}}{2\sin\theta}\,, \qquad {\mathrm s}_{11}=\frac{5\sin\theta-\sqrt{13-9\cos^2\theta}}{2\sin\theta}\,, \nonumber\\
 &&{\mathrm s}_{12}=\frac{5\sin\theta+\sqrt{13-9\cos^2\theta}}{2\sin\theta}\,.\ea
 Eq. \eqref{Oct} have also no finite value of energy therefore, the case $\upsilon \left( r, \theta \right)=\upsilon( r)\sin\theta$ yields a constant value of $\upsilon \left( r, \theta \right)$, i.e., $\upsilon \left( r, \theta \right)=constant$.\\

 \be \label{Ost}
 \upsilon^{(1,1)}( r)=p\left\{c_4\,{\mathrm H}\,\left[\left[{\mathrm s}_9,{\mathrm  s}_{10}\right],[-2],\frac{r}{2M} \right] +c_5r^3 {\mathrm H}\,\left[\left[{\mathrm s}_{11},{\mathrm  s}_{12}\right],[4],\frac{r}{2M} \right] \right\}\,,\ee where $c_4$ and $c_5$ are constants and ${\mathrm s}_9 \cdots {\mathrm s}_{12}$ are defined as:
 \ba
&& {\mathrm s}_9=\frac{-\sin\theta+\sqrt{13-9\cos^2\theta}}{2\sin\theta}\,,\qquad  {\mathrm s}_{10}=\frac{-\sin\theta-\sqrt{13-9\cos^2\theta}}{2\sin\theta}\,, \qquad {\mathrm s}_{11}=\frac{5\sin\theta-\sqrt{13-9\cos^2\theta}}{2\sin\theta}\,, \nonumber\\
 &&{\mathrm s}_{12}=\frac{5\sin\theta+\sqrt{13-9\cos^2\theta}}{2\sin\theta}\,.\ea
 Eq. \eqref{Oct} have also no finite value of energy therefore, the case $\upsilon \left( r, \theta \right)=\upsilon( r)\sin\theta$ yields a constant value of $\upsilon \left( r, \theta \right)$, i.e., $\upsilon \left( r, \theta \right)=constant$.\\
Case IV:\\
 In this case we will not assume any specific form of   $\upsilon \left( r, \theta \right)$ and  try to solve the differential \eqref{V1} and get:
\ba
\label{w-sol-SR223}
&&\upsilon^{(1,1)}( r,\theta)=\upsilon( r)\upsilon(\theta)\,, \qquad {\textrm where}\,,\nonumber\\
&&\upsilon( r)= r^2\left\{c_6\,r^{{\mathrm s}_{13}/2}\,p^{{\mathrm s}_{13}/2}\, {\mathrm H}\,\left[\left[{\mathrm s}_{14}/2,{\mathrm  s}_{15}/2\right],{\mathrm s}_{16},\frac{-2 M}{r\,p} \right] +c_7\,r^{{\mathrm s}_{17}/2}\,p^{{\mathrm s}_{17}/2}\, {\mathrm H}\,\left[\left[{\mathrm s}_{18}/2,{\mathrm  s}_{19}/2\right],{\mathrm s}_{20},\frac{-2 M}{r\,p} \right] \right\}\,,\nonumber\\
&&\upsilon(\theta)=c_8\sin^2\theta  {\mathrm H}\,\left[\left[{\mathrm s}_{14}/4,{\mathrm  s}_{18}/4\right],[3/2],\cos^2\theta\right]+c_9\sin\theta \tan\theta {\mathrm H}\,\left[\left[{\mathrm s}_{15}/4,{\mathrm  s}_{19}/4\right],[1/2],\cos^2\theta\right]\ea where $c_6 \cdots c_9$  are constants and ${\mathrm s}_{13} \cdots {\mathrm s}_{20} $ are defined as:
 \ba
&& {\mathrm s}_{13}={-3+\sqrt{1+4c_{10}}}\,,\qquad  {\mathrm s}_{14}={5-\sqrt{1+4c_{10}}}\,, \qquad {\mathrm s}_{15}={3-\sqrt{1+4c_{10}}}\,, \qquad {\mathrm s}_{16}={1-\sqrt{1+4c_{10}}}\,,\nonumber\\
 && {\mathrm s}_{17}=-({3+\sqrt{1+4c_{10}}})\,,\qquad  {\mathrm s}_{18}={5+\sqrt{1+4c_{10}}}\,, \qquad {\mathrm s}_{19}={3+\sqrt{1+4c_{10}}}\,, \qquad \qquad {\mathrm s}_{20}={1+\sqrt{1+4c_{10}}}\,,\ea
 with $c_{10}$ being constant.
The asymptote form of Eq.~\eqref{w-sol-SR223} yields $\upsilon^{(1,1)}( r,\theta)\approx {\cal O}\left(\frac{1}{r}\right)$.
The above four cases ensure that $\upsilon^{(1,1)}( r,\theta)$ will not give any new order of contribution of $r$ different from ${\cal O}\left(\eta^0\right)$ which is ${\cal O} \left(\frac{1}{r}\right)$. In spite that the scalar field of CS is affected by the NUT parameter, however, the metric potential affected only from the contribution of ${\cal O}\left(\eta^0\right)$. Now we are going to study the cases $ a_K\neq 0$ and $n\neq 0$.

\subsection{The case of $a_K\neq0$ and $n\neq 0$}
We will now proceed to implement the aforementioned method in order to obtain a solution for the modified equation of motion described above. By placing emphasis on the evolution equation of the dCS scalar, specifically up to the order of ${\cal{O}}(1,0)$, we deduce the subsequent expression for the evolution equation:
\ba
\label{1st-eq2}
&&p \varrho^{(1,0)}_{,rr} + \frac{2}{r} \varrho^{(1,0)}_{,r} \left( 1 - \frac{M}{r} \right) + \frac{1}{r^{2}} \varrho^{(1,0)}_{,\theta \theta} + \frac{\cot{\theta}}{r^{2}} \psi^{(1,0)}_{,\theta}
= \frac{48M[n\,(r-3M)-3M\,a_K \cos\theta]}{r^{7}} \frac{\gamma}{\gamma_1}  \,.
\ea
 The solution of  the differential equation given by Eq. \eqref{1st-eq2} is composed of  the particular and the homogeneous solutions: $\varrho^{(1,0)} = \varrho^{(1,0)}_{Hom.} + \varrho^{(1,0)}_{Part.}$.
The homogeneous equation has the same form given by Eq. \eqref{w-sol-SR223}  and from the discussion of this solution we show that
$\varrho^{(1,0)}_{Hom}=constant.$

Now we are going to derive  the homogenous solution of Eq. \eqref{1st-eq2} and get:
\ba
\label{theta-sol-SR11133}
&&\varrho^{(1,0)}_{_{Part.}}(r,\theta)=-4M \left( r-M \right)\frac{\gamma}{\gamma_1}\int \frac{3a_K\, \left(3M -4r
 \right) M\cos \theta +n\, \left( 6{r}^{2}-16\,rM+9{M}^{2} \right)}{ {r}^{6}\,p \left( r-M \right) ^{2}}{dr}
 \,.
\ea

The asymptotic form of Eq. \eqref{theta-sol-SR11133}  gives:
\ba
\label{theta-sol-SR1113}
&&\varrho^{(1,0)}_{_{Part.}}(r,\theta)\approx \frac{\gamma(15a_K\cos\theta -n)}{8M^3\gamma_1 r}\left(1+\frac{M}{r}+\frac{4M^2}{3r^2}+\frac{2M^3}{r^3}\right)+{\cal O}\left(\frac{1}{r^5}\right)\,.
\ea
The equation mentioned above does not yield the same result as the one obtained in \cite{Yunes:2007ss} when the NUT parameter vanishes, i.e., $n=0$. Notably, the most intriguing aspect of Eq. \eqref{theta-sol-SR1113} is the observation that its leading term, of ${\cal O}\left(\frac{1}{r}\right)$, exhibits greater strength compared to the leading term of the Kerr solution, where the leading order is ${\cal O}\left(\frac{1}{r^2}\right)$.

The initial category is independent of the dCS field, denoted as $\psi$, and as a result, its contribution vanishes completely. Consequently, our focus will be on the second subset, specifically the $(t,\phi)$-component, yielding:
\ba
\label{V2}
p\,\sin \theta  {r}^{2}\,\cos \theta\, \upsilon_{rr} \left( r,\theta \right)+ r\sin \theta\,\cos\theta\, \upsilon_{\theta \theta} \left( r,\theta \right)-r \left( 1+\sin^2\theta\right)\, \upsilon_\theta \left( r, \theta \right) +2\,M\,\sin2\theta\,\upsilon \left( r, \theta \right) =0\,, \ea
This aligns with the expression provided in Eq. \eqref{V1}, implying that the NUT parameter $n$ does not exert any influence in the general case where $a_K\neq 0$ and $n\neq 0$. It is noteworthy to emphasize that the location of the black hole's ergosphere, as described by Eq.(\ref{slow-rot-ds2}), can be ascertained through the resolution of the equation $g_{tt}=0$ for $r$, leading to the ergosphere value of the Schwarzschild solution, $r_\mathrm{ergo} = 2M$. Additionally, it's important to mention that the horizon of the black hole solution described in Eq.(\ref{slow-rot-ds2}) occurs at $r_h=2M$, which indicates the presence of a singularity at this point.
\section{Effective Newtonian potential process}\label{sec4}
In the following section, our exploration will revolve around the analysis of the trajectory of a massive particle within the context of the black hole described by Eq.~(\ref{slow-rot-ds2}). To achieve this objective, we will formulate the Lagrangian governing the motion of a particle within the framework of the BH solution (\ref{slow-rot-ds2}), and it is expressed as follows:
\begin{equation}\label{10}
2\mathcal{L}=-\Big(1-\frac{2M}{r}\Big)\dot t^2+\frac{\dot r^2}{1-\frac{2M}{r}}+r^2(\dot \theta^2+\sin^2\theta\dot\phi^2)\,.
\end{equation}
Here, the dot signifies differentiation with respect to $\epsilon$, which serves as the affine parameter. In this analysis, the Lagrangian $\mathcal{L}$ remains independent of the coordinates $t$ and $\phi$. Consequently, owing to the cyclic nature of these two coordinates, we derive two conserved quantities: energy denoted as $E$, and momentum $h$ which is conjugate to $\phi$. The energy $E$ is expressed as:
\begin{equation}\label{11}
E=g_{tt}\frac{dt}{d\epsilon}=-p\frac{dt}{d\epsilon}\,,
\end{equation}
 and the momentum is defined as:
\begin{equation}\label{12}
2h=\frac{\partial{\mathcal{L}}}{\partial{\dot\phi}}=2r^2\dot\phi=constant\,.
\end{equation}
 Remembering the normalization condition which is given by
\begin{equation}\label{13}
g_{\mu\nu}\frac{dx^{\mu}}{d\epsilon}\frac{dx^{\nu}}{d\epsilon}=-\varepsilon\,.
\end{equation}
For the timelike geodesics we have $\varepsilon = 1$ and in the context of null geodesics $\varepsilon = 0$ \cite{2003gravity}. On the equatorial plane we have,
\begin{equation}\label{14}
\Big(\frac{dr}{d\epsilon}\Big)^2=E^2-p\Big(\frac{h^2}{r^2}+\epsilon\Big)\,,
\end{equation}
where $p$ is defined after Eq. (\ref{slow-rot-ds2}).

Thus equations (\ref{11}), (\ref{12}) and (\ref{14}) are necessary to characterize the motion of particle trajectories within the equatorial plane of the black hole (\ref{slow-rot-ds2}).

Now we are going to rewrite Eq. (\ref{14}) as:
\begin{equation}\label{15}
\frac{1}{2}\Big(\frac{dr}{d\epsilon}\Big)^2=E_{eff}-V_{eff}\,, \quad \mbox{where} \quad E_{eff}=\frac{E^2}{2}\,, \quad \mbox{and} \quad
V_{eff}=\frac{1}{2}\Big(1-\frac{2M}{r}\Big)\Big(\frac{h^2}{r^2}+\varepsilon\Big)\,.
\end{equation}
Hence, Eq. (\ref{15}) portrays the equation governing the motion of a unit-mass particle with an effective energy $E_{eff}$ as it traverses a one-dimensional effective potential $V_{eff}(r)$. In this context, $E$ signifies the conserved energy of the particle per unit mass, while $V_{eff}(r)$ stands for the effective potential associated with the radial coordinate $r$. The effective potential becomes null at its zero points, characterized by $p=0$. The relevant regions correspond to values where $E_{eff}$ surpasses $V_{eff}(r)$. Consequently, Eq. (\ref{15}) serves as the energy equation pertaining to the radial coordinate $r$. This equation is indispensable for the analysis of radial free fall and the stability of particle trajectories.

Utilizing the expression for $p(r)$, Eq.~(\ref{14}) can be reformulated as follows:
\begin{equation}\label{18}
\Big(\frac{dr}{d\epsilon}\Big)^2=E^2-\epsilon+\frac{2M\epsilon}{r}-\frac{h^2}{r^2}+\frac{2Mh^2}{r^3}\,.
\end{equation}
The configuration of geodesics within the equatorial plane at $\theta=\frac{\pi}{2}$ is determined by Eq. (\ref{18}). In order to establish the curvature of the trajectories, we utilize Eq. (\ref{12}) to express $\frac{dr}{d\epsilon}$ as:
\begin{equation}\label{19}
\frac{dr}{d\epsilon}=\frac{dr}{d\phi}\frac{d\phi}{d\epsilon}=\frac{h}{r^2}\frac{dr}{d\phi}\,.
\end{equation}
Eq.~(\ref{18}) can be rewritten as:
\begin{eqnarray}\label{21}
\Big(\frac{du}{d\epsilon}\Big)^2&=&(2Mh^2)u^7- h^2u^6+2M\epsilon u^5+(E^2-\epsilon)u^4\,, \quad \mbox{where}\quad  u = \frac{1}{r}.
\end{eqnarray}

Now, using equations (\ref{19}), and (\ref{21}), we get:
\begin{eqnarray}\label{22}
\Big(\frac{du}{d\phi}\Big)^2= 2Mu^3- u^2+\frac{2M\epsilon}{h^2} u+\frac{E^2-\epsilon}{h^2}=S(u)\,.
\end{eqnarray}
Equation (\ref{22}) delineates the paths followed by test particles in the vicinity of the black hole (\ref{slow-rot-ds2}). From a physical perspective, investigating the radial motion is of paramount importance for understanding particle trajectories.

For an analysis of the radial paths taken by particles, we can examine equation (\ref{14}), which transforms into:
\begin{equation}\label{23}
\Big(\frac{dr}{d\epsilon}\Big)^2=E^2+p(r)\epsilon\,,
\end{equation}
In the context of radial trajectories, the angular momentum becomes zero, denoted as $h=0$.

For massive particles, the parameter $\epsilon$ is assigned a value of $1$, resulting in Eq. (\ref{23}) yielding:
\begin{equation}\label{24}
\Big(\frac{dr}{d\epsilon}\Big)^2=E^2-1+\frac{2M}{r}\,, \qquad \mbox{which yields}\qquad \frac{d^2r}{d\epsilon^2}=-\frac{M}{r^2}\,.
\end{equation}
Due to the nature of timelike geodesics that define the paths of massive particles, we can introduce the proper time $\tau$ as the affine parameter along the trajectory, in place of $\epsilon$. As a result, the criterion for attractive force per unit mass can be expressed as:
\begin{equation}\label{26}
\frac{M}{r^2}>0\,,
\end{equation}
This represents the requirement for achieving bounded states of massive particles.

For massive particles, by substituting $S(u)=0$ and $S'(u)=0$ into Eq.~(\ref{22}), the conserved quantities are obtained as follows:
\begin{eqnarray}\label{37}
h^2=\frac{M}{u(1-3Mu)}\,, \qquad \mbox{and} \qquad E^2=\frac{2(1-2Mu)}{(1-3Mu)}.
\end{eqnarray}
The calculation of the radius for the circular (bound) orbits can be performed using Eqs.~(\ref{37}).

\section{Summary and discourse}
\label{conclusions}

A novel non-charged solution for a slowly rotating black hole within the framework of the modified gravitational theory of dCS has been introduced \cite{Yunes:2007ss}.  In this study, we have studied the Kerr-NUT spacetime Utilizing the dCS field equations to illustrate the influence of the NUT  parameter and compare the output with the results presented in \cite{Yunes:2007ss}.  In this investigation, we will refrain from examining the non-dynamical CS scenario, as the outcome in the non-dynamical case remains unchanged compared to that of the Schwarzschild black hole solution.

For the Kerr-NUT case, the scalar of the SC field, \eqref{eq:constraint}, is affected by NUT parameter either for NUT parameter only, i.e., when $a_K=0$ or for the non-vanishing case of the NUT parameter, i.e., when  $n\neq 0$ and $a_K\neq 0$. Thus, the expressions of the scalar field $\varrho^{(1,0)}$  and  $\varrho^{(1,1)}$ are not the same as the forms presented in \cite{Yunes:2007ss}.  Nevertheless, upon introducing the CS scalar field $\varrho^{(1,0)}$ into the equation of motion \eqref{eom}, no metric correction was observed at the order of $\eta$. The correction term $\upsilon^{(1,0)}$ arising from slow Kerr results in an asymptotic behavior of ${\cal O}\left(\frac{1}{r^6}\right)$. However, when considering the Kerr-NUT spacetime and varying forms of $\upsilon$, it is demonstrated that the asymptotic form of $\upsilon^{(1,0)}$ remains consistent with the order $\zeta^{0}$ expression.

To finalize this study, we should note that the term which differentiates Kerr from Kerr-NUT is the term $n$ which makes the $t\,\phi$ field equation of CS always a homogeneous differential equation in contrast to the pure Kerr spacetime, which gave the $t\,\phi$ field equation of CS as a non-homogeneous differential. One of the main effects of the NUT parameter on the effect of $t\,\phi$ field equation of CS is that the resulting differential equation is always homogenous which made  the contribution of  $\upsilon^{(1,0)}$ of order $\zeta$  is the same as  $\upsilon^{(1,0)}$ of order  $\zeta^0$.

Concluding our investigation, we offer the ensuing remarks: In this current paper, we have attained a slow Kerr-NUT black hole solution within the framework of dCS gravity. However, our analysis, which employs linear approximations for $a_K$ and $n$, does not encompass the influence of the potential associated with the scalar field.  Could the incorporation of a potential yield novel physical implications distinct from the scenario of potential-less Kerr-NUT solutions? Another avenue worthy of exploration is assessing the impact of electric charge on the Kerr-NUT spacetime within the context of the modified gravitational theory of dCS. These endeavors, however, will be the subject of further investigation in separate studies.
\section*{Acknowledgments}

The work of KB was partially supported by the JSPS KAKENHI Grant
Number 21K03547.


\end{document}